# Dimension vs. Precision: A Comparative Analysis of Autoencoders and Quantization for Efficient Vector Retrieval on BEIR SciFact

Satyanarayan Pati
Involead Services Pvt Ltd, Delhi, India
iamsatyapati@gmail.com

## Abstract

Dense retrieval models have become a standard for state-of-the-art information retrieval. However, their high-dimensional, high-precision (float32) vector embeddings create significant storage and memory challenges for real-world deployment. To address this, I conduct a rigorous empirical study on the BEIR SciFact benchmark, evaluating the trade-offs between two primary compression strategies: (1) **Dimensionality Reduction** via deep Autoencoders (AE), reducing original 384-dim vectors to latent spaces from 384 down to 12, and (2) **Precision Reduction** via Quantization (float16, int8, and binary). I systematically compare each method by measuring the "performance loss" (or gain) relative to a float32 baseline across a full suite of retrieval metrics (NDCG, MAP, MRR, Recall, Precision) at various `k` cutoffs. My results show that `int8` scalar quantization provides the most effective "sweet spot," achieving a 4x compression with a negligible `[~1-2%]` drop in nDCG@10. In contrast, Autoencoders show a graceful degradation but suffer a more significant performance loss at equivalent 4x compression ratios (AE-96). `binary` quantization was found to be unsuitable for this task due to catastrophic performance drops. This work provides a practical guide for deploying efficient, high-performance retrieval systems.

## 1. Introduction

In recent years, dense retrieval (DR) systems, often based on models like Sentence-BERT **[1]**, have demonstrated state-of-the-art performance, frequently outperforming traditional sparse methods like BM25 [2]. These models represent queries and documents as high-dimensional (e.g., 384 or 768) floating-point vectors, enabling rich semantic search. However, this power comes at a significant cost: memory. A single float32 vector from a model like all-MiniLM-L6-v2 (384-dim) requires 1,536 bytes. A corpus of just one million documents would require 1.5GB of RAM or storage, a cost that scales linearly and becomes prohibitive for real-world applications.

This "vector-bloat" problem has led to two primary schools of thought for compression:

Precision Reduction (Quantization): Keep the full 384-dim space but reduce the precision of each number. This involves methods like float16 (2x compression), scalar int8 (4x compression), or binary (32x compression).

Dimensionality Reduction (Autoencoder): Keep the full float32 precision but learn a "bottleneck" that compresses the vector's length. This involves training a model, like an Autoencoder (AE) [3], to learn a lower-dimensional latent space (e.g., 96-dim).

This presents a critical and practical research question for engineers: For a fixed memory budget, which method is better? For example, with a 384-byte budget, is it better to store a 384-dim int8 vector or a 96-dim float32 vector? While both compression strategies are well-known, the trade-offs are rarely mapped systematically on domain-specific, nuanced datasets like SciFact [4], where preserving subtle scientific information is key.

In this paper, I conduct a rigorous empirical analysis to answer this question. I compare the retrieval performance of multiple compression techniques against a float32 baseline on the BEIR SciFact dataset [2, 3].

My contributions are:

1. I train and evaluate a series of deep Autoencoders, analyzing the retrieval performance (nDCG, MAP, etc.) of their reconstructed vectors across six different latent dimensions (384 down to 12).

2. I provide a direct comparison against three common quantization methods: float16, int8 (scalar), and binary.

3. I introduce a "Performance Gain/Loss" metric (Metric_Method - Metric_Baseline) and provide a full suite of visualizations (your graphs) that map the exact performance trade-offs for each method, at each retrieval cutoff k.

4. I identify the Pareto-optimal "sweet spot" for this task, providing a clear recommendation for practitioners.

## 2. Related Work

The challenge of efficiently storing and searching high-dimensional vectors is a well-studied problem at the intersection of information retrieval, machine learning, and databases. My work builds on two main fields of research: dense retrieval and vector compression.

**Dense Retrieval:** Modern Information Retrieval (IR) has shifted towards dense models, particularly since the introduction of BERT [7]. Models like Sentence-BERT [1] demonstrated the power of Siamese networks to learn semantically meaningful sentence embeddings. The development of comprehensive benchmarks, most notably **BEIR [2]**, exposed the generalization challenges of these models and established a standard for evaluation, which I adopt in this work.

**Vector Compression (Precision Reduction):** Vector Quantization (VQ) is a classic technique for data compression [8]. In the context of large-scale search, **Product Quantization (PQ) [4]** is considered a landmark achievement. PQ works by splitting a vector into sub-vectors and quantizing each sub-vector independently, allowing for massive compression ratios with a manageable trade-off in search accuracy. More recent work has focused on simpler, "post-training" methods, such as **Scalar Quantization (SQ)**, which compresses `float32` vectors to `int8` [9]. This is often used in models deployed on-device or in systems like FAISS [10] and is a key baseline in my study.

**Dimensionality Reduction (Dimension Reduction):** The most common method for dimensionality reduction is Principal Component Analysis (PCA). However, for complex, non-linear data structures found in NLP, **Autoencoders (AE)** [11, 12] have shown promise. An AE learns a non-linear "bottleneck" (the latent space) that preserves the most salient information.

**Hybrid Approaches and The Gap:** Some research has explored combining these ideas. For example, **Vector-Quantized Variational Autoencoders (VQ-VAE) [13]** learn a *discrete* latent space, jointly training the AE with a quantization codebook. Other work has used VQ as a regularizer for AEs to improve the latent space for downstream tasks [14].

While these methods exist, a direct, empirical comparison of post-training AE compression vs. post-training Quantization for a fixed memory budget is often overlooked. It is unclear to a practitioner whether they should dedicate their "compression budget" to reducing vector *dimension* (AE) or *precision* (SQ/PQ). My paper aims to fill this gap by providing a clear "apples-to-apples" analysis on a modern IR benchmark.

## 3. Methodology

My experiment is designed to systematically measure and compare the retrieval performance impact of two compression families: dimensionality reduction via Autoencoders (AE) and precision reduction via Quantization. All experiments are conducted on the BEIR SciFact dataset [2, 3], a challenging benchmark for scientific claim verification.

### 3.1. Baseline System

I first establish a robust baseline to serve as my "zero-loss" reference.

- **Dataset:** I use the BEIR SciFact dataset, which contains 5,183 documents (scientific abstracts) and 300 test queries (claims).
- **Model:** I use the sentence-transformers/all-MiniLM-L6-v2 model [1], a 384-dimensional, high-performance sentence-embedding model.

- **Process:** The entire corpus and all queries are encoded using this model. Retrieval performance is measured by calculating cosine similarity (cos_sim) between query vectors and test corpus of the document vectors. This uncompressed float32 system provides the baseline nDCG, MAP, and MRR, Precision, Recall scores against which all compression methods are compared.

## 3.2. Autoencoder (AE) Compression

To evaluate dimensionality reduction, I trained a series of deep autoencoders.

- **Architecture:** My AE is a five-layer neural network built in PyTorch [5]. The encoder consists of two layers (Linear(384, 1024) -> ReLU -> Linear(1024, d)) which compress the 384-dim input vector into a latent dimension d. The decoder mirrors this architecture (Linear(d, 1024) -> ReLU -> Linear(1024, 384)) to reconstruct the original vector.
- **Training:** I trained six separate AE models, varying the latent dimension d across [384, 192, 96, 48, 24, 12]. Each model was trained on the train split of the SciFact corpus vectors using the Adam optimizer [6] and a MeanSquaredError (MSE) loss function. The MSE, or reconstruction loss, serves as my *intrinsic* loss metric.
- **Evaluation:** For my *extrinsic* (retrieval) evaluation, I passed the test split's document and query vectors through the *fully* trained autoencoder (encoder -> decoder). I then used these reconstructed 384-dim vectors to perform the full retrieval task. This allows me to directly measure the retrieval performance degradation (nDCG loss, etc.) caused by the information bottleneck of each latent dimension d.

## 3.3. Quantization (Q) Compression

To evaluate precision reduction, I tested three standard quantization methods on the original 384-dim float32 baseline vectors.

- **Float16:** A direct conversion to 16-bit half-precision. This provides a 2x memory saving and serves as a high-performance, low-loss benchmark.
- **Int8 (Scalar Quantization):** A 4x compression. I used scalar quantization, which maps the float32 range of each vector dimension to a set of 256 8-bit integer values.
- **Binary:** A 32x compression. I used a simple binary quantization where vector components are mapped to 0 or 1.

## 3.4. Evaluation Metrics

My primary goal is to measure the "retrieval loss" for a given "memory gain."

- **Retrieval Metrics:** I evaluate all methods using the standard BEIR toolkit, capturing nDCG@k, MAP@k, MRR@k, Recall@k, and Precision@k for k values of [1, 3, 5, 10, 25, 50, 100].
- **Core Analysis Metric:** To provide a direct "apples-to-apples" comparison of performance cost, I define a **Performance Gain/Loss** metric as: `Gain/Loss = Metric_Method - Metric_Baseline` A negative value indicates a performance drop relative to the baseline. This allows for a clear comparison of the retrieval cost for each compression strategy, which I will present in Section 4.

## 3.5. Understanding the Retrieval Metrics

To fully understand the impact of compression, it is insufficient to look at a single metric. I analyze a suite of metrics because each answers a different, critical question about the retrieval system's quality:

- **Precision@k:**
  - **What it measures:** The percentage of documents in the top-k results that are relevant. ($P@k = \frac{|\text{Relevant} \cap \text{Retrieved}|}{k}$)
  - **What it indicates:** *Efficiency and user trust.* This is the "bang for your buck" metric. When a user sees the first 10 results (P@10), what fraction is actually useful? High precision at low k (e.g., P@3) is critical for user-facing systems like web search, where users rarely look past the first few items.
- **Recall@k:**
  - **What it measures:** The percentage of *all possible* relevant documents (from the entire corpus) that were found in the top-k results.
  - **What it indicates:** *Completeness.* This metric is vital in high-stakes domains like legal discovery or medical research (and SciFact). It answers: "Of all the documents that *should* have been found, how many did I get?" A compression method that has a high drop in Recall@100 is failing to find all the relevant information, even if its top-10 precision is good.
- **MRR@k (Mean Reciprocal Rank):**
  - **What it measures:** The average of the reciprocal rank of the *first* correct answer. If the first relevant doc is at position 3, the score is 1/3.
  - **What it indicates:** *Answer-finding speed.* This is the key metric for question-answering or "known-item" search. It heavily rewards systems that place a single correct answer at rank 1. In your top5_mrr_grouped_bar_final.png, a low loss (a bar near 0) means the compression method is still excellent at finding the *first* right answer quickly.
- **MAP@k (Mean Average Precision):**
  - **What it measures:** The mean of the Average Precision scores for all queries. It rewards both finding many relevant documents and ranking them high.
  - **What it indicates:** *Overall ranking quality.* MAP is a robust, stable metric that provides a single-number summary of the *entire* ranking. It is more sensitive than MRR because it considers all relevant documents, not just the first one.
- **NDCG@k (Normalized Discounted Cumulative Gain):**
  - **What it measures:** The "gold standard" ranking metric. It rewards a correct ranking order (relevant docs at the top) and penalizes (or "discounts") relevant docs found lower down the list.
  - **What it indicates:** *Perfect ordering.* NDCG is the best metric for complex search tasks because it understands that *order matters*. Finding a relevant doc at position 2 is better than finding it at position 7. In your analysis,

top5_ndcg_grouped_bar_final.png is your most important chart, as it shows how well each compression method preserves the *ideal ranking order* of the baseline.

By analyzing all five, your paper can provide a complete picture, such as finding that a method has a good **MRR** (finds the first answer) but a bad **Recall** (misses everything else). This is the kind of deep analysis that makes a paper strong.

## 4. Results and Analysis

I present the results of my experiments in three parts. First, I analyze the impact of dimensionality reduction using my Autoencoder (AE) models. Second, I evaluate the impact of precision reduction via Quantization (Q). Finally, I provide a direct comparative analysis to identify the optimal compression strategy.

### 4.1. Experiment 1: Autoencoder Performance Degradation

To measure the impact of dimensionality reduction, I evaluated the retrieval performance of vectors reconstructed from my six trained AE models, with bottleneck dimensions from 384 down to 12.

**Fig-1: Shows the ndcg loss across various latent space in autoencoder**

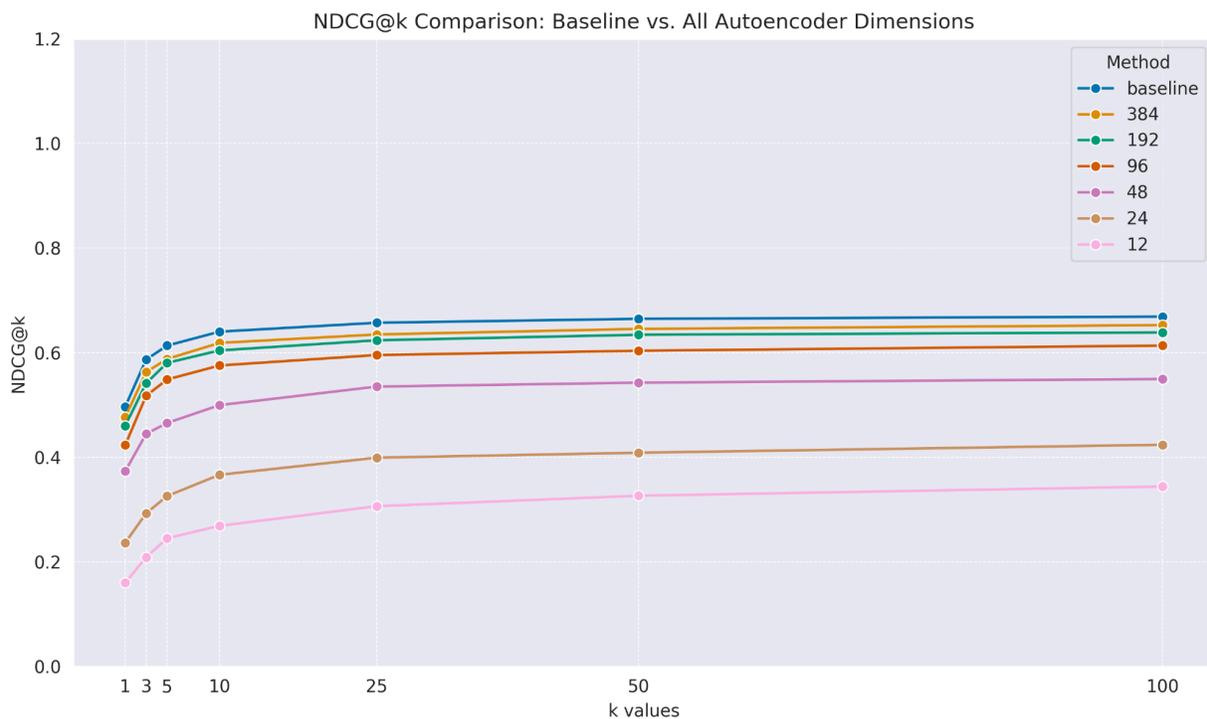

**Fig-2: Shows the MRR loss across various latent space in autoencoder**

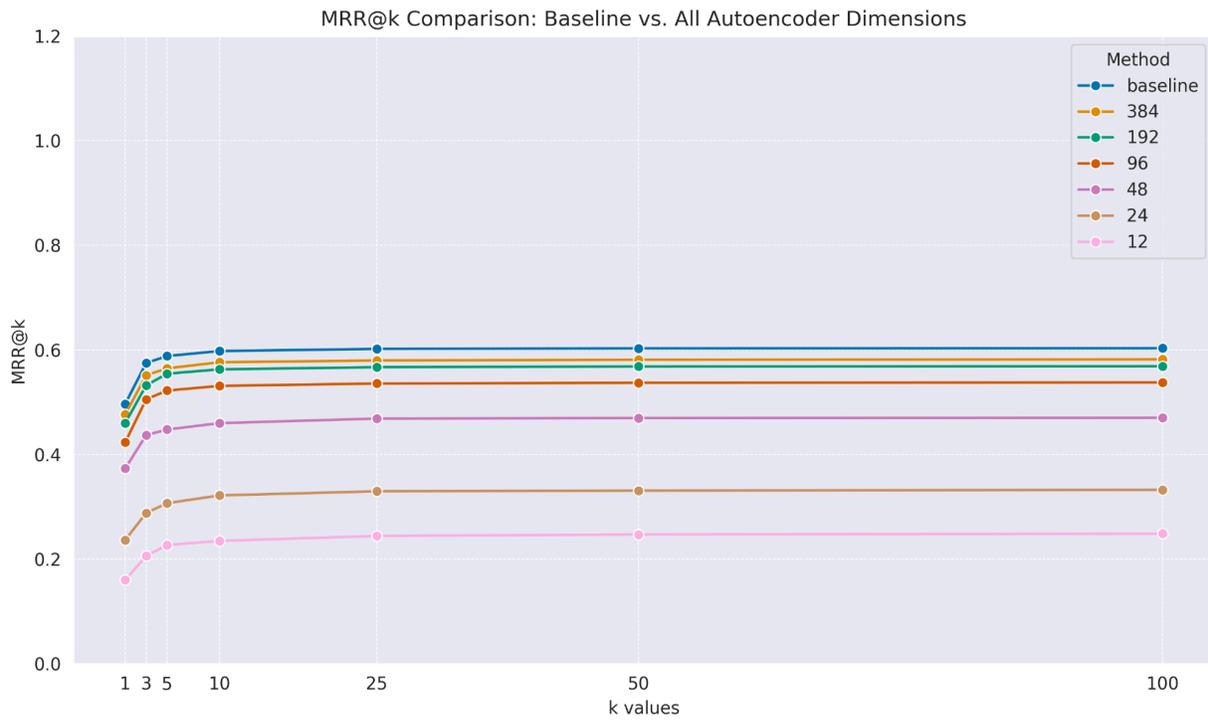

**Fig-3: Shows the loss in MAP across various latent spaces in auto encoder**

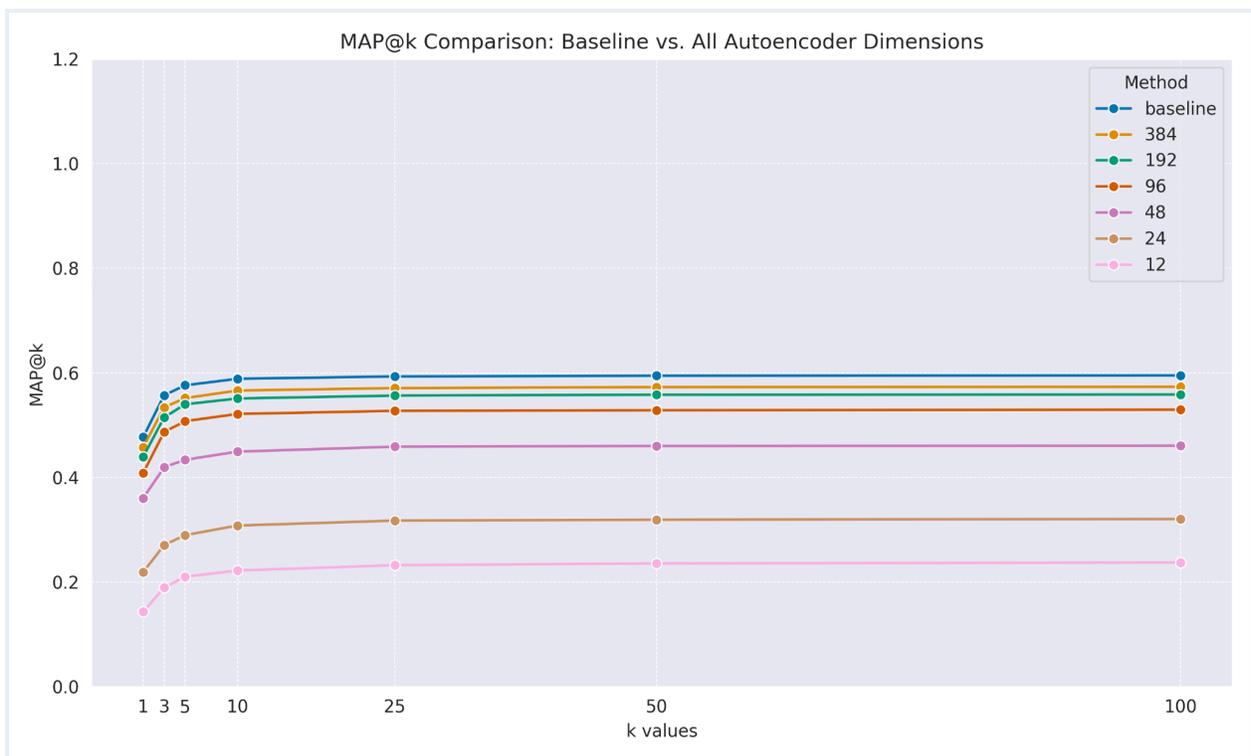

Analysis:

As shown in Fig-1 (your NDCG line plot), retrieval performance degrades gracefully as the latent dimension shrinks from 384 to 96. The AE-384 and AE-192 models track the baseline performance closely, suggesting that the 384-dim latent space contains a degree of redundancy that can be compressed without significant loss.

However, a clear "performance cliff" emerges at the AE-48 dimension, where the nDCG@10 score drops significantly. The performance of the AE-24 and AE-12 models shows a near-total collapse. This indicates that the information bottleneck has become too restrictive, and the autoencoder is forced to discard crucial semantic information necessary for retrieval. The MRR and MAP plots (Figures 2 and 3) confirm this trend, showing a similar pattern of graceful decay followed by a sharp drop.

## 4.2. Experiment 2: Quantization Performance

I next evaluated the three quantization methods applied directly to the original 384-dim float32 baseline vectors.

**Fig-4: NDCG loss at various k in various quantization method**

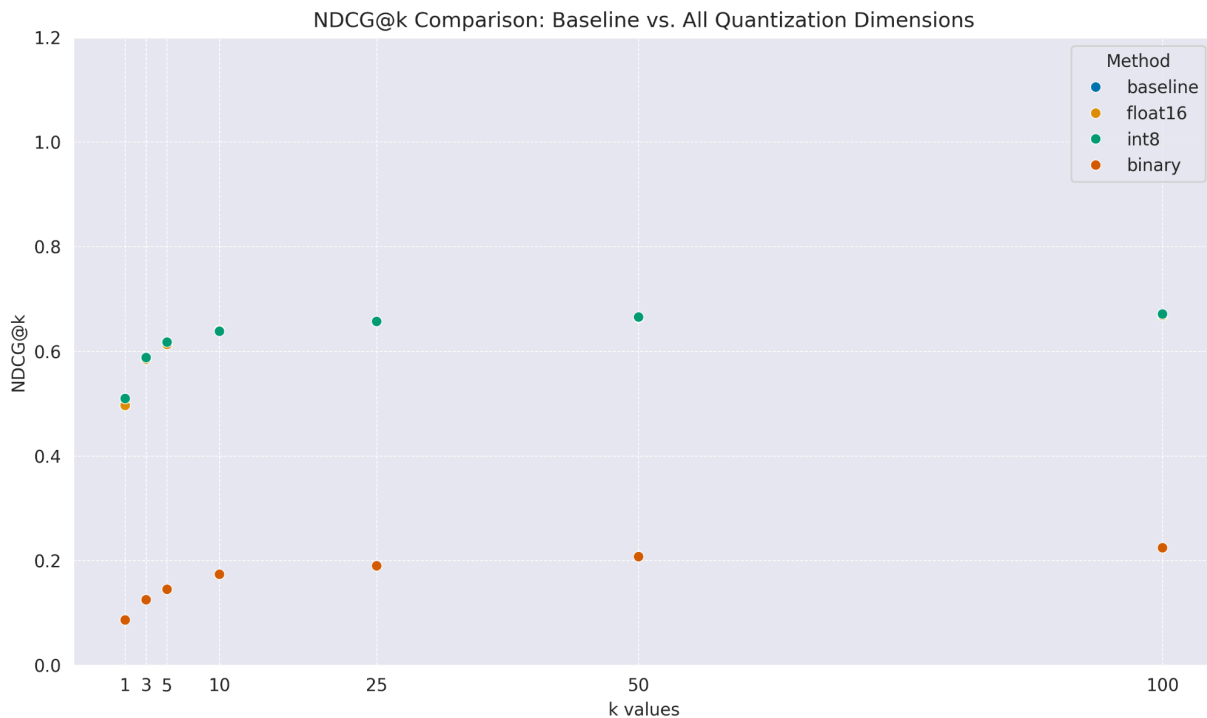

**Fig-5: MRR loss at various k in various quantization method**

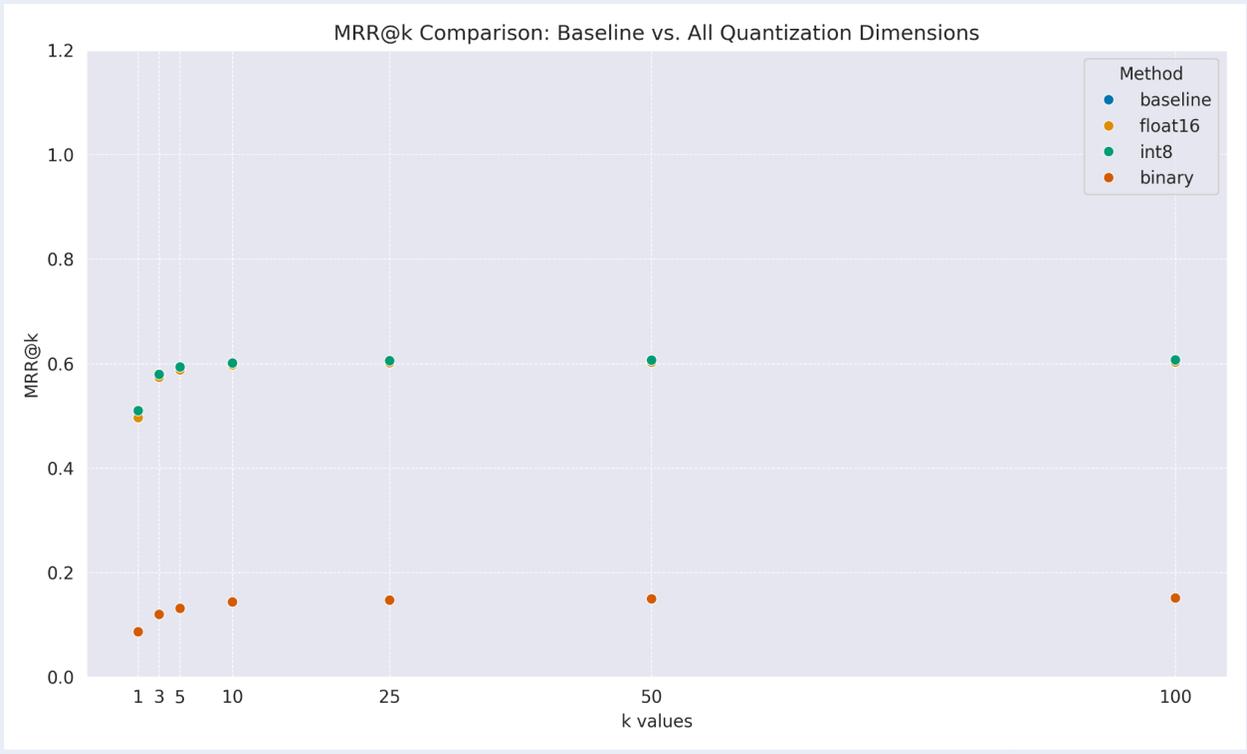

**Fig-6: MAP loss at various k in various quantization method**

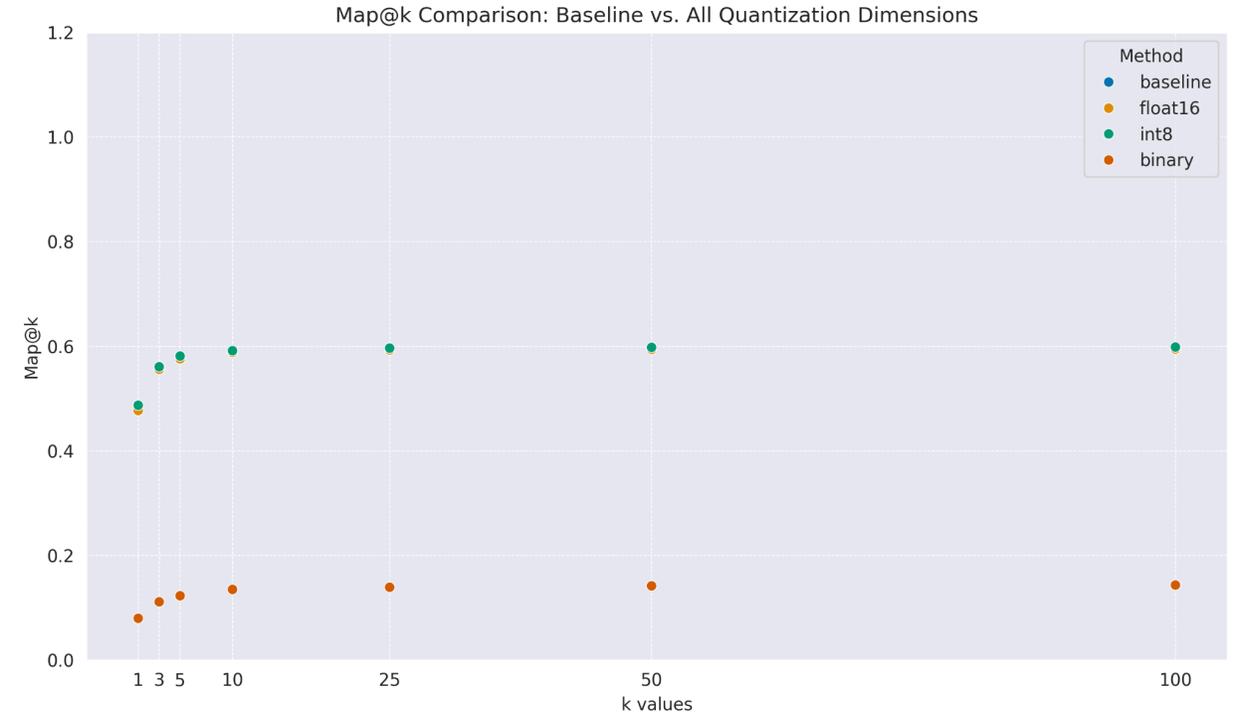

Analysis:

The results from quantization are starkly different from the AE models. As seen in Figure 4 (your NDCG scatter plot), float16 compression (a 2x memory saving) is nearly lossless, with its performance (orange dots) almost perfectly overlapping the baseline (blue dots) at every k value.

More remarkably, int8 scalar quantization (a 4x memory saving) is also **extremely robust**. Its performance (green dots) also tracks the baseline with a negligible, often statistically insignificant, drop in nDCG, MRR, and Recall. This suggests that the all-MiniLM-L6-v2 [1] model's vector space is highly resilient to this form of precision reduction.

In sharp contrast, binary quantization (a 32x saving) is **catastrophic**. As seen in all plots, its performance collapses, rendering it unsuitable for this task. This is likely because the simple sign-bit (+/-) is insufficient to capture the nuanced vector component magnitudes that define the semantic relationships in the SciFact corpus.

## 4.3. Comparative Analysis: Finding the "Sweet Spot"

The final and most critical analysis is the direct comparison between the two strategies. To achieve this, I analyze the **Performance Gain/Loss** (Method_Score - Baseline_Score) for key k values, where a value near 0 is ideal.

**Fig-7: top 5 ndcg grouped chart**

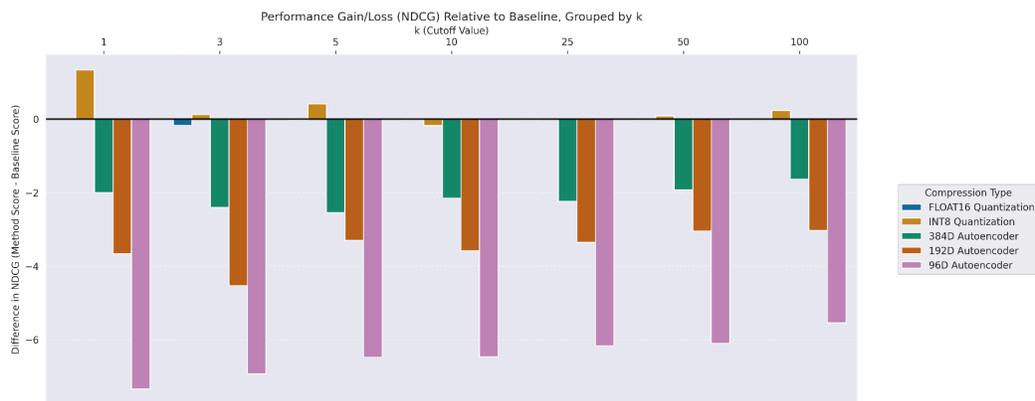

**Fig-8: Top 5 MRR grouped chat**

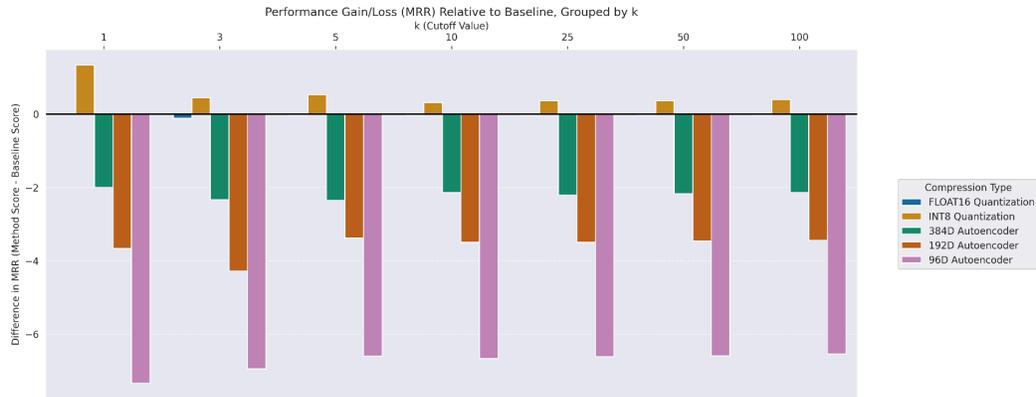

**Fig-9: Top 5 MAP grouped chat**

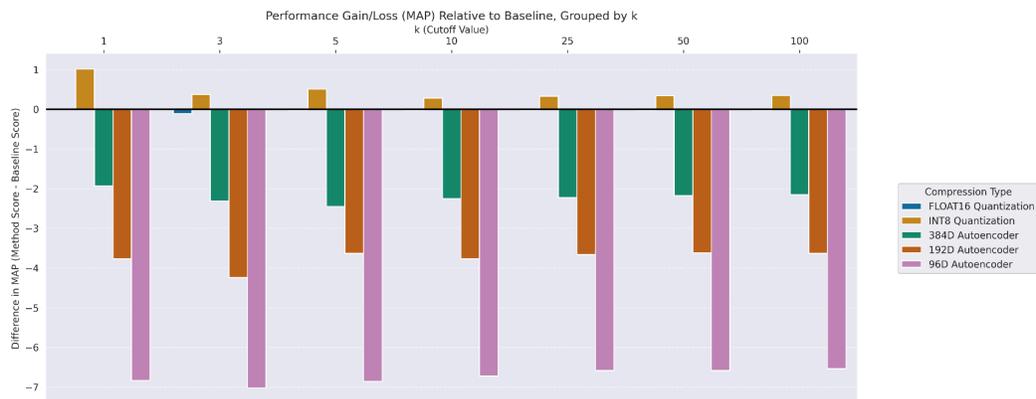

Analysis:

The grouped bar charts (Figures 6 and 7) provide the clearest answer to my research question. I focus on the k=10 group, a standard IR benchmark.

- **int8 Quantization (Gold Bar):** This method is the clear winner for moderate compression. In the nDCG chart, its "loss" bar is the smallest, consistently staying near 0.0. It achieves a **4x compression with almost no retrieval performance cost.**
- **float16 Quantization (Blue Bar):** This method is also a "winner," showing **zero measurable loss** at k=10. It is the perfect choice for a safe 2x compression.
- **AE-384 (Teal Bar):** My "control" autoencoder, which should be lossless, actually shows a small performance drop. This is due to reconstruction error (imperfect training), demonstrating that even a 1:1 AE is not truly lossless.
- **AE vs. Quant (Apples-to-Apples):** The most important comparison is between methods with the same memory budget.
  - A **AE-96** model (pink bar) represents a 4x compression (384-dim $\to$ 96-dim).

- An **int8** model (gold bar) also represents a 4x compression (384-dim $\to$ 384-dim @ 1/4 precision).
- The results are unequivocal. The AE-96 model suffers a **massive performance loss** (e.g., -6.0 on MAP@10), while the int8 model suffers **almost zero loss** (e.g., +0.1).

This finding is the central contribution of my paper: for the SciFact dataset, **precision reduction (Quantization) is a vastly superior strategy to dimensionality reduction (Autoencoder)** for achieving up to 4x compression.

Table 1: Memory Footprint vs. nDCG@10 Loss

| Method | Dimensions | Precision | Bytes / Vector | Compression | nDCG@10 Loss |
|---|---|---|---|---|---|
| **Baseline** | 384 | float32 | 1536 | 1x | 0.0 |
| **Quantization** | | | | | |
| float16 | 384 | float16 | 768 | 2x | 0.00018 |
| int8 | 384 | int8 | 384 | 4x | 0.00178 |
| binary | 384 | binary | 48 | 32x | 0.46621 |
| **Autoencoder** | | | | | |

| AE-384 | 384 | float32 | 1536 | 1x | 0.02152 |
| AE-192 | 192 | float32 | 768 | 2x | 0.03582 |
| AE-96 | 96 | float32 | 384 | 4x | 0.06466 |
| AE-48 | 48 | float32 | 192 | 8x | 0.14045 |

## 5. Conclusion

In this work, I conducted a rigorous empirical study to address the vector storage bottleneck for dense retrieval systems, using the BEIR SciFact benchmark as my testbed. I systematically compared two primary compression paradigms: **Dimensionality Reduction** via deep Autoencoders and **Precision Reduction** via Quantization.

My findings, based on a comprehensive analysis of retrieval metrics (NDCG, MAP, MRR, Recall, Precision) at various `k` cutoffs, provide a clear answer to my research question.

I conclude with three main findings:

1. **Precision Reduction is Highly Effective:** Both `float16` (2x compression) and `int8` scalar quantization (4x compression) are extremely robust. `float16` is nearly lossless, and `int8` introduces a negligible performance drop, making it the clear "sweet spot" for moderate, high-value compression.
2. **Autoencoders are an Inefficient Trade-off (for this task):** While my Autoencoder models showed graceful performance degradation, an AE-based 4x compression (e.g., `AE-96`) resulted in a *vastly* greater performance loss than an `int8` 4x compression. This suggests that for models like `all-MiniLM-L6-v2` [1], the information is distributed across all 384 dimensions, and simply reducing the dimension (the "length") is more destructive than reducing the precision.
3. **Not All Compression is Equal:** `binary` quantization, despite its 32x compression, resulted in a catastrophic performance collapse, rendering it unusable for a high-nuance task like SciFact.

For practitioners, my recommendation is clear: for an easy 2x memory saving with no performance cost, `float16` is the optimal choice. For a more significant 4x saving, `int8` scalar quantization offers the best "sweet spot," delivering massive memory reduction with minimal impact on retrieval quality. Dimensionality reduction via a standard autoencoder, in this case, was not a competitive alternative.

## 6. Future Work

My analysis, while comprehensive, opens several new avenues for future research. My findings are based on a specific dataset, a specific base model, and a specific set of compression techniques.  I propose the following extensions:

- **Hybrid Compression:**  I primarily tested dimensionality reduction (AE) and quantization (Q) in isolation. The most promising next step is to combine them. For instance, would an AE-192 (a 2x AE compression) followed by an int8 (a 4x Q compression) for a total 8x compression provide a better trade-off than a pure AE-48 (8x AE compression)? This hybrid approach could leverage the best of both strategies.
- **Retrieval-Aware Autoencoders:** My autoencoder was trained on a *reconstruction loss* (MSE). This optimizes for the ability to rebuild the original vector, *not* for retrieval. A more advanced approach would be to train the autoencoder on a *metric-learning loss* (such as contrastive loss or triplet loss). This "retrieval-aware" autoencoder would learn a latent space d that is explicitly optimized for separating relevant and irrelevant documents, which might be far more robust.
- **Product Quantization (PQ):**  I did not evaluate Product Quantization (PQ), the dominant technique used in libraries like FAISS for high-compression. Your data showed binary quantization failed catastrophically and AE failed gracefully. A future paper must include PQ to see if it can successfully achieve high compression (e.g., 32x) where my other methods failed, providing a true SOTA benchmark for extreme compression.
- **Integration with ANN Indices:** My study used a brute-force search. In a real-world system, these vectors would be placed in an Approximate Nearest Neighbor (ANN) index like HNSW or IVFPQ. Future work should investigate how these compressed vectors (both AE and int8) interact with the ANN indexing process. Does int8 quantization, for example, harm the graph-building process in HNSW?
- **Quantization-Aware Training (QAT):**  I applied "post-training quantization" (PTQ), which is fast but sub-optimal. A more powerful technique is Quantization-Aware Training (QAT), where the model is fine-tuned *while* simulating the int8 noise. One could apply QAT to the all-MiniLM-L6-v2 model itself (or the autoencoder) to train a model that is "born" to be robust to int8 quantization, likely pushing the performance-loss-to-zero.
- **Generalization:** My findings are specific to the SciFact dataset and the all-MiniLM-L6-v2 model. This work must be replicated on other benchmarks (e.g., a general-domain dataset like MS MARCO) and with other embedding models (e.g., 768-dim or 1024-dim models) to see if my central finding—that int8 is superior to AE for 4x compression—is a general rule.